\documentclass[conference]{IEEEtran}
\IEEEoverridecommandlockouts
\usepackage{cite}
\usepackage{geometry}

\AtBeginDocument{%
  \providecommand\BibTeX{{%
    Bib\TeX}}}
\usepackage{booktabs}
\usepackage{amsmath}
\usepackage{amssymb}
\usepackage{amsmath,amssymb,amsfonts}
\usepackage{algorithmic}
\usepackage{graphicx}
\usepackage{balance}
\usepackage{textcomp}
\usepackage{threeparttable}
\usepackage{xcolor}
\usepackage{subfigure} 
\def\BibTeX{{\rm B\kern-.05em{\sc i\kern-.025em b}\kern-.08em
    T\kern-.1667em\lower.7ex\hbox{E}\kern-.125emX}}
\begin{document}

\title{An Efficient Sparse Hardware Accelerator for Spike-Driven Transformer\\
\thanks{This work was supported in part by the Shenzhen Science and Technology
Program 2023A007.
Corresponding author:  Wendong Mao.}
}

\author{\IEEEauthorblockN{Zhengke Li$^a$, Wendong Mao$^a$, Siyu Zhang$^b$, Qiwei Dong$^b$ and Zhongfeng Wang$^a$$^,$$^b$}
\textit{$^a$School of Integrated Circuits, Sun Yat-Sen University, China}\\
\textit{$^b$School of Electronic Science and Engineering, Nanjing University, China}\\
Email: lizhk37@mail2.sysu.edu.cn, maowd@mail.sysu.edu.cn, syzhang@smail.nju.edu.cn, \\qiweidong@smail.nju.edu.cn, wangzf83@mail.sysu.edu.cn}

\maketitle

\begin{abstract}
Recently, large models, such as Vision Transformer and BERT, have garnered significant attention due to their exceptional performance. However, their extensive computational requirements lead to considerable power and hardware resource consumption. Brain-inspired computing, characterized by its spike-driven methods, has emerged as a promising approach for low-power hardware implementation. In this paper, we propose an efficient sparse hardware accelerator for Spike-driven Transformer. We first design a novel encoding method that encodes the position information of valid activations and skips non-spike values. This method enables us to use encoded spikes for executing the calculations of linear, maxpooling and spike-driven self-attention. Compared with the single spike input design of conventional SNN accelerators that primarily focus on convolution-based spiking computations, the specialized module for spike-driven self-attention is unique in its ability to handle dual spike inputs. By exclusively utilizing activated spikes, our design fully exploits the sparsity of Spike-driven Transformer, which diminishes redundant operations, lowers power consumption, and minimizes computational latency. Experimental results indicate that compared to existing SNNs accelerators, our design achieves up to 13.24$\times$ and  1.33$\times$ improvements in terms of throughput and energy efficiency, respectively.
\end{abstract}


\section{Introduction}
In recent decades, Artificial Neural Networks (ANNs), such as Convolutional Neural Networks (CNNs)~\cite{b2.5} and Transformer~\cite{WOS:000452649406008}, have achieved significant success in various fields. However, the continuous expansion of network structures has made ANNs with extensive parameters and computations challenging to implement on edge devices with limited hardware resources and power consumption. 

Spiking Neural Networks (SNNs)~\cite{b2} have emerged as a promising approach for energy-efficient implementation. By mimicking the neurodynamics of the human brain, SNNs use neurons that communicate between layers via spike-form signals and typically involve temporal accumulation operations. The binary nature of SNNs simplifies their operations, offering more potential directions for designing hardware accelerators tailored to SNNs.




Developing efficient hardware accelerators tailored to specific networks has been a prevalent approach to accelerating the inference of ANNs on edge devices~\cite{c1,c2,c3}. Although these designs optimize computations of convolution and attention through parallelism, sparsity, and data reuse, they are unable to efficiently exploit the characteristics of SNNs such as spiking data and timesteps in spatial and temporal dimensions. Thus, it is crucial to design dedicated accelerators for SNNs.

Currently, most existing SNN accelerators target driving types (time-driven, event-driven, or hybrid-driven), sparsity (reconfiguring computation kernels for different sparsity), data reshaping for spiking neurons, or data reuse with reconfigurable designs~\cite{b3,b4,b5,b6}. However, these accelerators only focus on the inherent characteristics of spiking neurons, without incorporating optimizations for specific layers like maxpooling and spike-driven self-attention (SDSA). Additionally, the dual spike inputs of self-attention are rarely considered in previous designs. To address this, we develop an efficient hardware accelerator for Spike-driven Transformer~\cite{b7} to fully exploit its high sparsity and optimize spiking computations, thereby reducing hardware costs with lower power consumption and processing latency.
The main contributions of this paper are as follows:
\begin{itemize}
\item We propose a novel spike redundancy elimination computation method to fully utilize the sparsity in the Spike-driven Transformer. This method encodes the position information of spikes and converts spiking computations into address comparison, which can effectively bypass zeros in the single and dual spike inputs.

\item Based on the proposed efficient computation method, we design low-complexity spike encoding and computing units to optimize multiple types of spiking computations, including maxpooling, SDSA, and linear. Benefiting from the binary nature of SNNs and our encoding method, all spiking calculations are implemented through simple addition and comparison instead of multiplication.

\item The proposed hardware accelerator is implemented on Xilinx Virtex UltarScale FPGA and evaluated with Spike-driven Transformer on the Cifar-10 dataset. Experimental results show that compared to other SNN accelerators, our design achieves up to 13.24$\times$ and 1.33$\times$ improvements in peak performance and energy efficiency, respectively.
\end{itemize}

\section{Preliminary of Spike Neuron Model}
Inspired by the way that neurons in the human brain process and transmit signals, SNNs have been proposed and considered as the third generation of neural networks~\cite{b2}. In SNNs, the Leaky Integrate-and-Fire (LIF) spiking neuron~\cite{b10} is generally used to encapsulate the main characteristics of biological neurons, whose dynamic behaviors can be summarized by the following equations:
\begin{flalign}
Temp[t]&=S[t]V_{reset}+(1-S[t])(\gamma Mem[t]),\label{1}
\end{flalign}
\begin{flalign}
Mem[t]&=Spa[t]+Temp[t-1],\label{2}
\end{flalign}
\begin{flalign}
S[t]&=\varepsilon(Mem[t]-V_{th}),\label{3}
\end{flalign}



{
\setlength{\parindent}{0cm}
where $Mem[t]$, $Spa[t]$, and $Temp[t]$ represent the membrane potential, spatial input, and temporal input of the neuron at timestep $t$, respectively. $S[t]$ is the output of the spiking neuron at timestep $t$. $\varepsilon(x)$ denotes the step function, whose output is `1' when $x$ is greater than or equal to 0; otherwise, the output is `0'. $V_{th}$ represents the threshold voltage required for the spiking neuron to fire a spike. $\gamma$ is the time constant representing the decay of the membrane potential. $V_{reset}$ represents the reset value of the membrane potential.
}

An intuitive understanding of spiking neuron's behavior is as follows: spatial input comes from various sources such as convolution, linear or maxpooling layers, while temporal input is determined by whether the spiking neuron fires a spike in the previous timestep. If the neuron fires, $Mem[t]$ resets to $V_{reset}$; otherwise, the membrane potential decays according to $\gamma$, which is typically between 0 and 1. From Equations (\ref{1})-(\ref{3}), it is evident that the computational mechanism of spiking neurons involves timestep, meaning whether a spike is fired depends on the current input $Spa[t]$ and the previous $Temp[t-1]$. 




\section{Hardware Architecture for Spiken-driven Transformer}

\begin{figure}[htbp]
  \centering
  \includegraphics[width=8.7cm,page={1}]{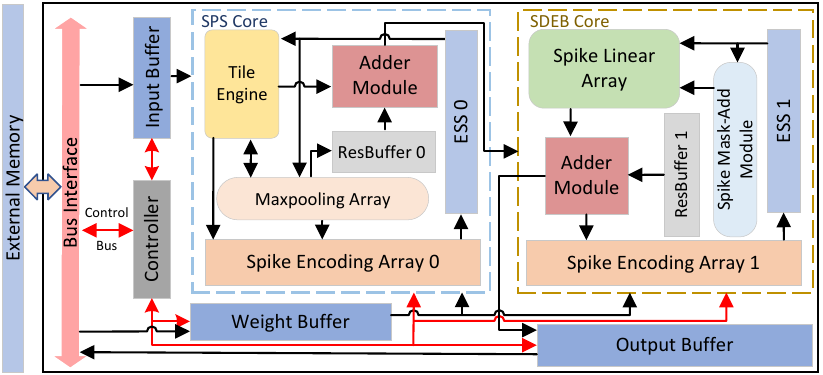}
  \caption{Overall architecture of our hardware accelerator.}
  \label{fig.1}
\end{figure}

Considering the characteristics of the Spike-Driven Transformer~\cite{b7}, our hardware accelerator focuses on optimizing operations with encoded spike input to fully utilize the sparsity of SNNs. The overall structure of our accelerator is shown in Fig.~\ref{fig.1}. The \textbf{Input Buffer} and \textbf{Output Buffer} interface with external data. The \textbf{Controller} generates signals to control each computational module and data interaction. The \textbf{SPS Core} and \textbf{SDEB Core} are designed to support the calculations of two key components in the Spike-Driven Transformer (i.e., Spiking Patch Splitting (SPS) and Spike-driven Encoder Block (SDEB)), respectively. Each core contains a \textbf{Spike Encoding Array (SEA)} and an \textbf{Encoded Spike SRAM (ESS)} to generate and store encoded spikes. The \textbf{Maxpooling Array} in the SPS Core integrates \textbf{Spike Maxpooling Units (SMUs)} and conventional maxpooling modules to perform maxpooling with spike or regular input. To support the calculation of SDSA, a \textbf{Spike Mask-Add Module (SMAM)} is developed, which can effectively handle dual spike inputs. \textbf{ResBuffer} and \textbf{Adder Module} are employed for residual computations.

External data is first transmitted to the Input Buffer, and then sent to the \textbf{Tile Engine} \cite{b12.5} in SPS Core to perform convolution computations. After residual computation, the output data of the SPS Core is sent to the \textbf{SEA} in the SDEB Core for encoding. The encoded spikes are processed through the SMAM and the \textbf{Spike Linear Array (SLA)} to complete the computations of SDSA and multilayer perceptron (MLP). Finally, the results are written back to the Output Buffer and passed to the External Memory.

\subsection{Spike Encoding Unit}

\begin{figure}[htbp]
  \centering
  \includegraphics[width=.45\textwidth,page={1}]{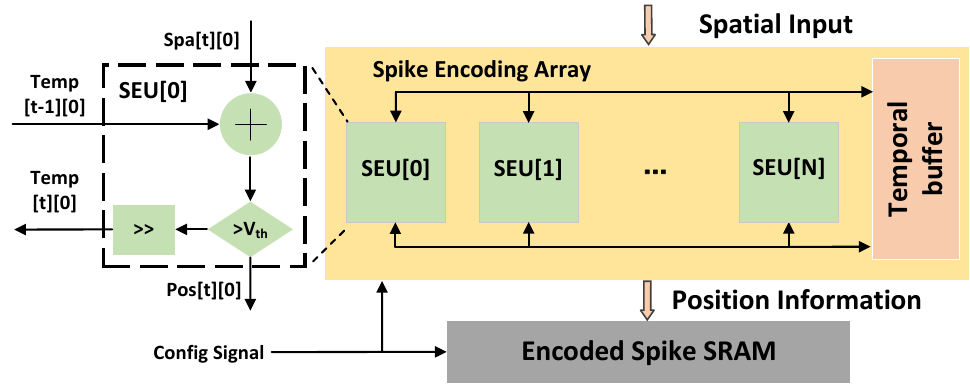}
  \caption{The architecture of the proposed SEA. When the output of the adder exceeds the firing threshold $V_{th}$, the current address $Pos[t][0]$ is stored in the ESS. $Temp[t][0]$ and $Spa[t][0]$ represent temporal and spatial input for SEU[0] in timestep $t$.}
  \label{fig.2}
\end{figure}

The proposed SEA comprises multiple Spike Encoding Units (SEUs) that embed position information into spikes. For example, given a matrix $\mathbf{I} \in \mathbb{R}^{C \times H \times W}$, we first reshape it to $\mathbf{I'} \in \mathbb{R}^{C \times L}$, where $L=H\times W$ represents the number of tokens. When a spiking neuron fires a spike, the token position information $Pos$ of the spike is encoded as output to replace the original binary values and stored in ESS, as shown in Fig.~\ref{fig.2}.

Although encoding position information requires additional logic and storage resources, the encoding format is efficient when passing the spikes to subsequent modules for processing. 
Compared to the bitmap encoding scheme, the modules do not need to determine whether it is a spike before calculation, thereby reducing the time required for indexing valid data and improving computational efficiency.
Encoded spikes are stored sequentially according to address order, which is crucial for designing subsequent computational units.

\subsection{Spike Maxpooling Unit}\label{AA}
\begin{figure}[htbp]
  \centering
  \includegraphics[width=8cm,page={1}]{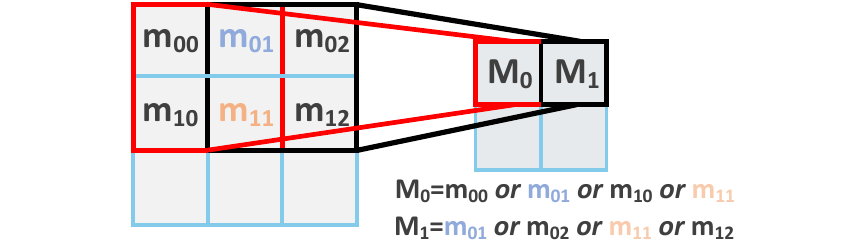}
  \caption{Illustration of the calculation process of the SMU. The red and black boxes represent the two adjacent kernel. ``$or$'' represents the logical OR.}
  \label{fig.3}
\end{figure}

An SMU is designed to optimize maxpooling with spike inputs, by using the encoded position information of spikes instead of comparing all values within a kernel to determine the output. As shown in Fig.~\ref{fig.3}, we set the kernel size to 2$\times$2 and stride of 1 as an example to illustrate the calculation process of the proposed SMU. Due to the binary output of spiking neurons, as long as the kernel covers the position of spikes, its output is `1'. 
If there is horizontal and vertical overlap between kernels, the overlapping data is reused to determine the output of multiple kernels simultaneously. For example, in Fig.~\ref{fig.3}, when the encoded spike is $m_{01}$, the values of both $M_{0}$ and $M_{1}$ are `1'. With the sparsity brought by spiking neurons, the SMU only captures spikes for calculation, saving redundant operations in performing maxpooling. 
 
\begin{figure}
\begin{center}
\subfigure[]{
\includegraphics[width=8cm]{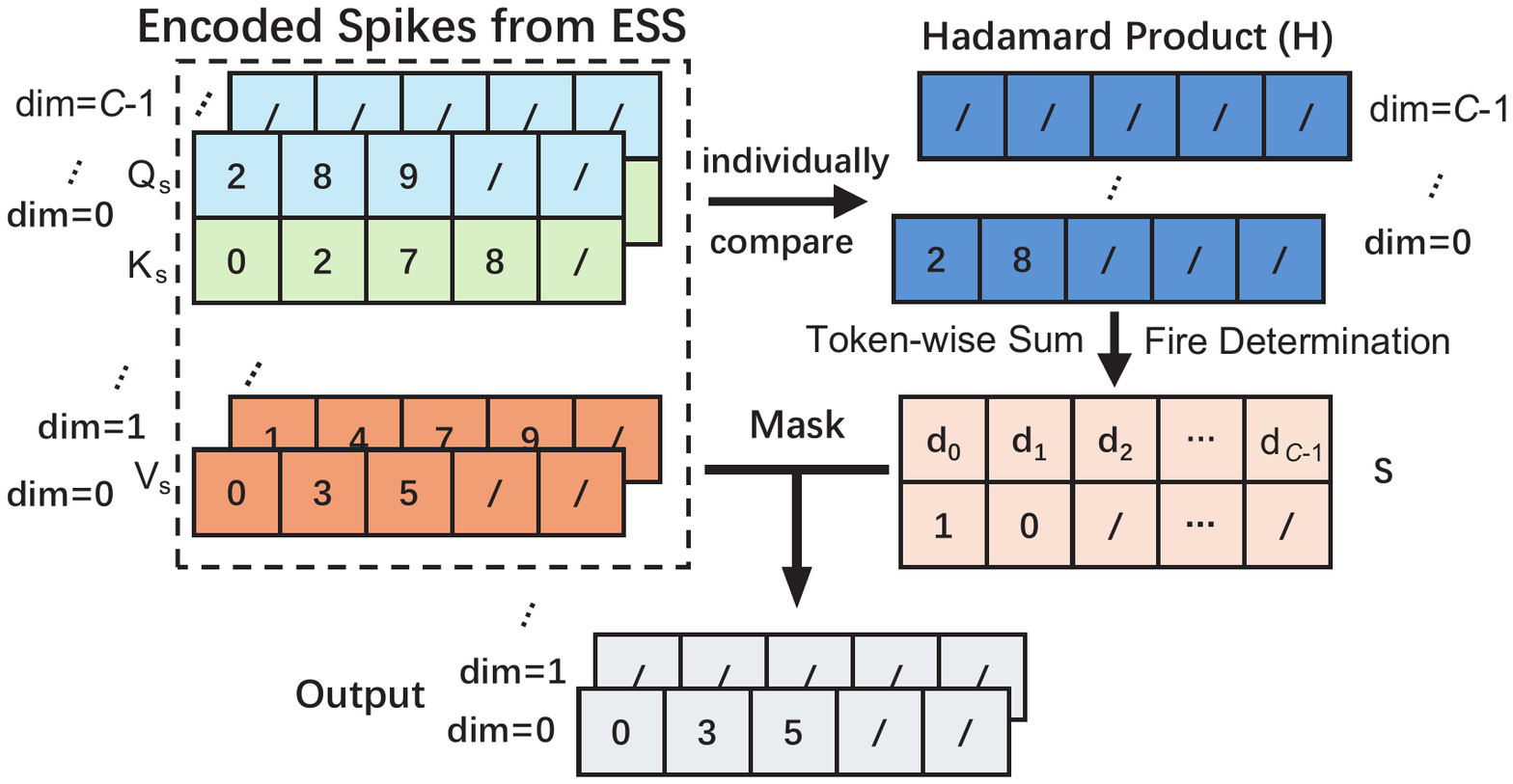}
}
\subfigure[]{
\includegraphics[width=4.5cm]{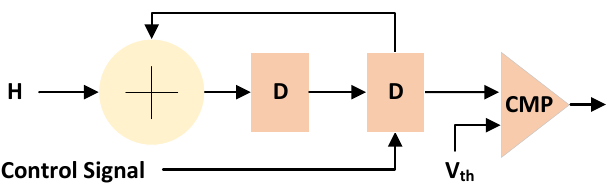}
}
\subfigure[]{
\includegraphics[width=3.5cm]{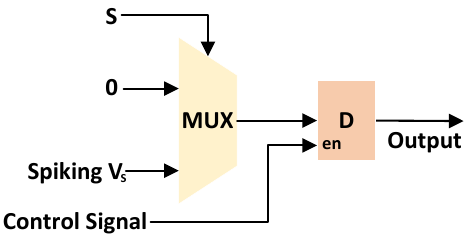}
}
\caption{The details of SMAM. (a) Data paths of SMAM. (b) The logic of token-wise accumulation and fire determination. (c) The logic of masking.}
\label{fig.4}
\end{center}
\end{figure}

\subsection{Spike Mask-Add Module}

In the Spike-Driven Transformer~\cite{b7}, complex attention calculations in the original Transformer~\cite{WOS:000452649406008} are replaced by Mask-Add operations. The multiplication of \textbf{Q} and \textbf{K} matrices is converted into the Hadamard product of two binary spike matrices ($\mathbf{Q_{s}}$ and $\mathbf{K_{s}}$), whose result is then accumulated along the token dimension. The accumulated result is converted into a binary value through spiking neurons and used as a mask for spiking $\mathbf{V_{s}}$. After masking, the data passes through linear and Batch Normalization (BN) to obtain the output of SDSA.

Fig.~\ref{fig.4} shows the details of our SMAM, which implements the main computations in SDSA with dual spike inputs. The specific calculation process is as follows:


\begin{itemize}
\item First, the position information of spiking $\mathbf{Q_{s}}$ and $\mathbf{K_{s}}$ is encoded and stored separately, as shown in the Fig.~\ref{fig.4}(a). One encoded spike is taken into the comparator, followed by another from the other memory, which differs from the data temporarily stored in the comparator. 
\item Next, if the two encoded spikes have the same address, the comparator outputs `1'; otherwise, the comparator temporarily stores the larger address and repeats the data-taking and comparison until all encoded spikes of $\mathbf{Q_{s}}$ or $\mathbf{K_{s}}$ have been compared. The result of the Hadamard product is accumulated along the token dimension.
\item Finally, the result of accumulation is compared with the firing threshold $V_{th}$ to generate the mask $\mathbf{S}$. If the values exceed the firing threshold, the corresponding position in $\mathbf{S}$ is set to `1'; otherwise, it is set to `0'. The specificity of the mask allows the channel storing the encoded spikes of $\mathbf{V_{s}}$ to be cleared or retain its values based on $\mathbf{S}$.
\end{itemize}

\begin{figure}
\begin{center}
\subfigure[]{
\includegraphics[width=7cm]{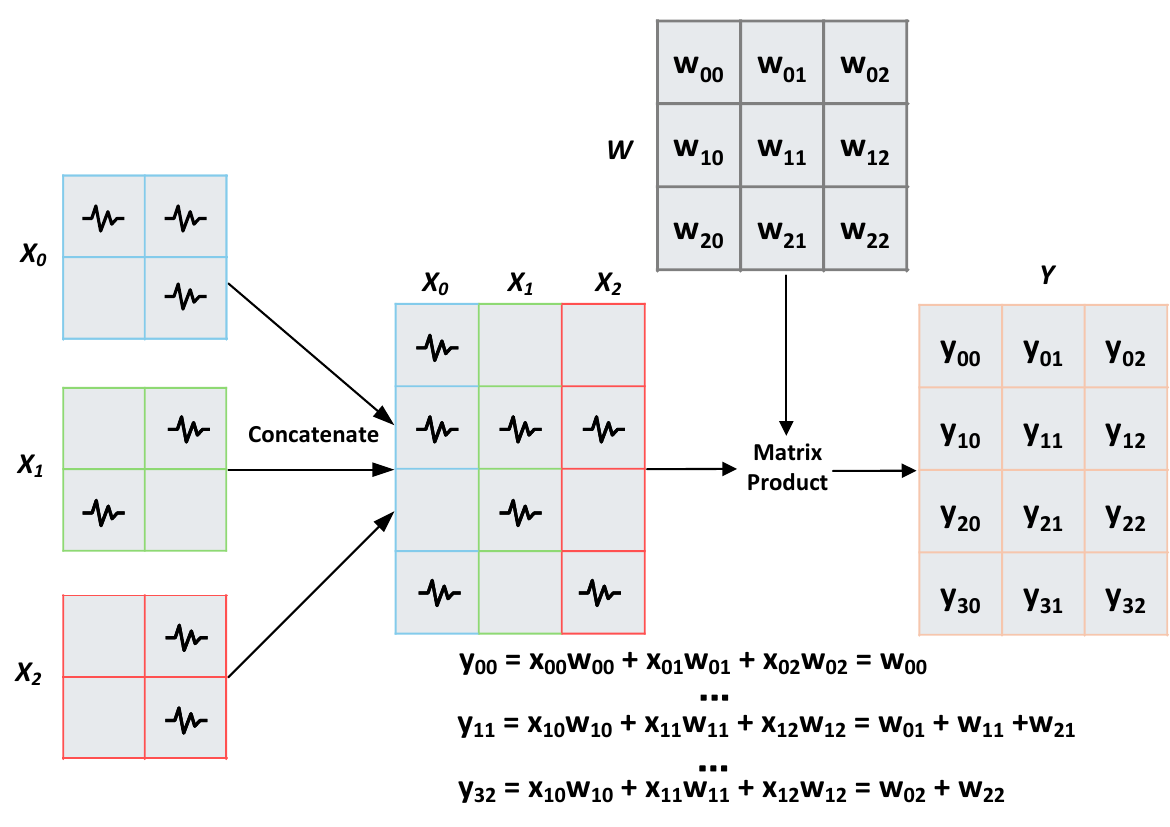}
}
\subfigure[]{
\includegraphics[width=.4\textwidth]{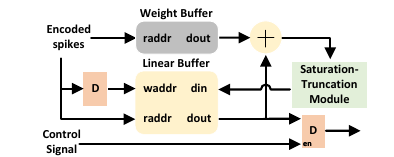}
}
\caption{Illustration of the SLU. (a) The calculation process of the SLU with an input matrix $\mathbf{X}\in \mathbb{R}^{3 \times 2 \times 2}$. (b) The architecture of the SLU. The Saturation-Truncation Module prevents the value from wrapping around to the negative side or the positive side, ensuring the result fits within the specified bit width.}
\label{fig.5}
\end{center}
\end{figure}

\subsection{Spike Linear Unit}

We design a Spike Linear Unit (SLU) to perform linear computations with spike inputs, which is multiplication-free. The input matrix $\mathbf{X}\in \mathbb{R}^{C \times H \times W}$ of the SLU is also flattened to $\mathbf{X'} \in \mathbb{R}^{C \times L}$. Since the input is in spike form, operations in the SLU involve only sparse addition, as shown in Fig.~\ref{fig.5}. Based on the encoded position information of spikes, weights in the corresponding positions are selected for accumulation. By skipping zero values in the spike inputs, the SLU enhances the computational efficiency of the linear layer, reducing processing latency and power consumption. Additionally, since encoded spikes are stored in different memory banks based on their channels, the input channel can serve as a parallel extension to further accelerate computation.

\section{Experiment and Results}

\subsection{Experimental Setup}

\begin{figure}[htbp]
  \centering
  \includegraphics[width=7cm]{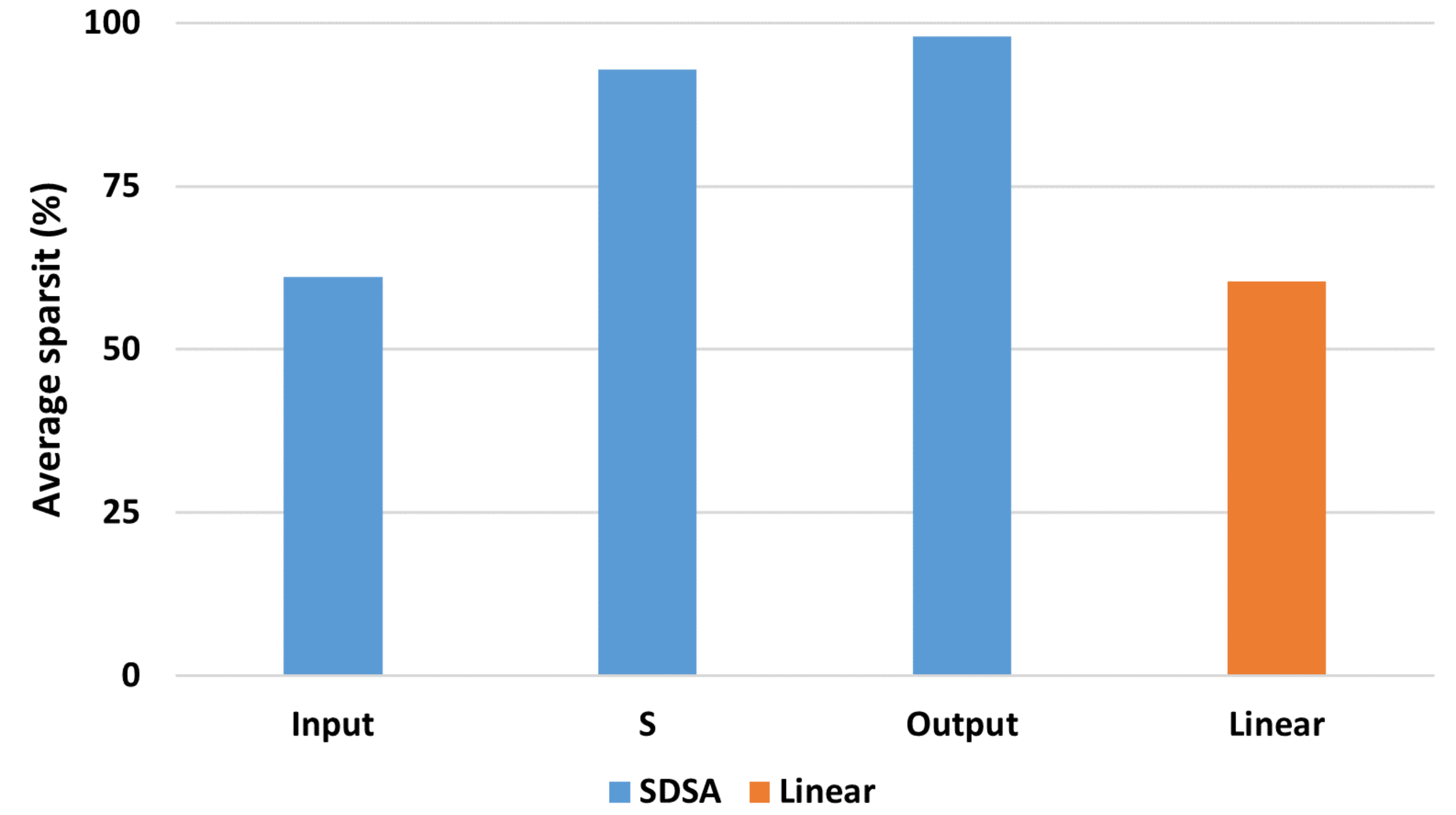}
  \caption{The average sparsity of SDSA and subsequent linear layers.}
  \label{fig.6}
  \vspace{-0.1cm}
\end{figure}

To verify the feasibility of the spike encoding method and evaluate its performance, we selected the Spike-driven Transformer \cite{b7} as the benchmark for our accelerator. Fig.~\ref{fig.6} illustrates the average sparsity of several modules in~\cite{b7}. We applied 10-bit quantization for weights and activations, and 8-bit quantization for encoded spikes. After quantization, the model achieved 94.87\% accuracy on the CIFAR-10 dataset. 

The designed accelerator has been fully implemented on a Xilinx Virtex UltraScale FPGA using Verilog HDL for coding and Vivado 2023.2 for hardware utilization and power estimation. It can simultaneously compute the outputs of up to 1,536 spiking neurons, with a clock frequency of 200 MHz.


\subsection{Hardware Implemental Results}

         

Table~\ref{table.2} presents a comparison of our SNN accelerator with others. ``Synaptic operation" (SOP)  refers to a spike traversing a unique synapse. By fully leveraging the sparsity of SNNs and applying our encoding method to compute maxpooling, linear, and self-attention operations, our accelerator achieves higher peak performance (up to $13.24\times$) and more energy-efficient (up to $1.33\times$) implementation compared to other SNN accelerators. Additionally, it is worth noting that, compared to other SNN accelerators that only adapt spikes to CNN or FC, our accelerator can support more complex computations with spike inputs in Transformer networks. To achieve this, our design incorporates a more complex storage mechanism, which incurs additional memory resource costs for storing temporal data at each timestep and extra logic to encode and process spikes. Despite these costs, the results show that our design achieves excellent performance, with a peak throughput of 307.2 GSOP/s and an energy efficiency of 25.6 GSOP/W.  

\begin{table}
  \caption{Comparison with Other SNN Accelerators}
  \label{table.2}
  \begin{threeparttable}\scalebox{0.94}{
  \begin{tabular}{ccccc}
    \toprule
    & ISCAS \cite{b15} &   TCAD\cite{b13}&AICAS\cite{b17}&Ours\\ 
    \midrule
         Year&  2022& 2022&2023 &2024\\ 
         Network&  FC&   CNN&CNN&Trans.*\\ 
         Dataset& MNIST&   MNIST&MLND&Cifar-10\\ 
         
         Platform&  Kintex Ultra.&Zynq7000  &Zynq Ultra.&Virtex Ultra.\\ 
         LUT& 416296 &   45986&41930&453266\\  
         FF&  95000& 20544 &16237&94120\\ 
         BRAM&  216**&  262 &128&784\\ 
         Freq.(MHz)&   140 & 200&200&200\\ 
         GSOP/s&  179**&  22.6 &23.2&\textbf{307.2}\\ 
         GSOP/W&  21.49**& 19.3  &19.3& \textbf{25.6}\\ 
    \bottomrule
  \end{tabular}}
  \begin{tablenotes}
        \footnotesize
        \item *Transformer with spike inputs.
        \item **Take the average performance under different conditions.
      \end{tablenotes}
  \end{threeparttable}
\end{table}

\section{Conclusion}

This paper proposes a novel processing method for the Spike-driven Transformer that encodes the position information of spikes to effectively bypass non-spike values. Based on this method, we design efficient computational units: SMU, SLU, and SMAM. Experimental results demonstrate that the developed accelerator achieves a peak throughput of 307.2 GSOP/s and an energy efficiency of 25.6 GSOP/W, representing improvements of 13.24$\times$ and 1.33$\times$ compared to other SNN accelerators. Furthermore, unlike most SNN accelerators that concentrate on the linear layer, our design also applies the encoded spikes in maxpooling and SDSA with dual spike inputs. Combined with our proposed design, SNNs can be more effectively and efficiently mapped to resource-constrained devices, achieving faster inference in practical scenarios.


\end{document}